\pdfoutput=1
\documentclass{JINST}
\let\ifpdf\relax
\usepackage{amsmath}

\title{Polarisation and Beam Energy Measurement at a Linear $\mathrm{e}^+\mathrm{e}^-$ Collider}

\author{B.~Vormwald$^{a}$\\
\llap{$^a$}Deutsches Elektronen-Synchrotron DESY\\
           Notkestr. 85 \\
           22607 Hamburg, Germany\\
           E-mail: \email{benedikt.vormwald@desy.de}}

\abstract{The International Linear Collider (ILC) is a future electron/positron collider at
the energy frontier. Its physics goals are clearly focused on precision measurements 
at the electroweak scale and beyond. Beam energy and beam polarisation
are two important beam parameters, which need to be measured and monitored
to any possible precision. We discuss in this publication the foreseen concepts of
beam energy and beam polarisation measurement at the ILC:

Two Compton polarimeters per beam line will determine the beam polarisation.
The anticipated precision of this measurement amounts to $\Delta \mathcal{P} / \mathcal{P} =2.5 \times 10^{-3}$,
which is a challenging goal putting highest demands on detector alignment and
linearity. Recent detector developments as well as a detector
calibration technique are described, which allow for meeting these requirements.
The beam energy is measured before and after the interaction point to a
targeted precision of $\Delta E/E = 10^{-4}$. Thereby, the two foreseen concepts are introduced:
A noninvasive energy spectrometer based on beam position monitors
is planned to be operated before the interaction region. Behind, a synchrotron
radiation imaging detector will allow not only for measuring the beam energy,
but also gives access to the beam energy spread of the (disrupted) beam.}

\keywords{ILC; polarimetry; beam energy measurement}

\newcommand{\reftitle}[1]{``#1'',}

\newcommand{\MGy}{\mathrm{\,MGy}}

\newcommand{\GeV}{\mathrm{\,GeV}}
\newcommand{\TeV}{\mathrm{\,TeV}}
\newcommand{\km}{\mathrm{\,km}}
\newcommand{\m}{\mathrm{\,m}}
\newcommand{\nm}{\mathrm{\,nm}}
\newcommand{\cm}{\mathrm{\,cm}}
\newcommand{\mm}{\mathrm{\,mm}}
\newcommand{\mum}{\mathrm{\,\mu m}}

\newcommand{\MHz}{\mathrm{\,MHz}}

\newcommand{\ms}{\mathrm{\,ms}}

\newcommand{\murad}{\mathrm{\,\mu rad}}
\newcommand{\mrad}{\mathrm{\,mrad}}

\newcommand{\h}{\mathrm{\,h}}

\newcommand{\sqrts}{\sqrt{s}}

\newcommand{\electron}{e}

\newcommand{\muon}{\mu}

\newcommand{\approptoinn}[2]{\mathrel{\vcenter{
  \offinterlineskip\halign{\hfil$##$\cr
    #1\propto\cr\noalign{\kern2pt}#1\sim\cr\noalign{\kern-2pt}}}}}

\begin{document}

\section{Introduction}
The International Linear Collider (ILC) is a future $31\km$ long linear $e^+/e^-$ collider operating at the energy frontier.
In its baseline design, it is foreseen to be operated at a center-of-mass energy of $\sqrts=500\GeV$ with an optional upgrade to $\sqrts=1\TeV$.
ILC is going to provide longitudinally polarised electron/positron beams.
Thereby, the beam polarisation is defined like
\begin{equation}
\mathcal{P}_z=\frac{N_R-N_L}{N_R+N_L},
\end{equation}
where $N_{R/L}$ describes the number of left/right-handed polarised electrons/positrons in the particle beam.
At the ILC, an electron beam polarisation of at least $\mathcal{P}_z=80\%$ and a positron polarisation of at least $30\%$ is planned.
Recently, the ILC Technical Design Report has been published~\cite{tdr}.

The ILC physics program is clearly aiming for high precision measurements.
This requires not only excellent detectors, but also high-performance beam line instrumentation.
The collision of elementary particles offers the unique possibility of a precise knowledge of the initial state of an event like the center-of-mass energy and beam polarisation.
In the following, we will present the beam-energy and polarisation measurement concepts at the ILC.

The main energy spectrometer and polarimeter are situated in the beam delivery system (BDS), which is the last about $2\km$ of the ILC beam line before the interaction point (IP)  providing among the beam diagnostics also the beam collimation and the final focus to the nanometer scale.
Within the BDS, the polarimeter is located about $1700\m$ and the main energy spectrometer $700\m$ upstream the IP.

At the ILC, also an instrumented extraction line is foreseen comprising a second energy spectrometer as well as second polarimeter situated $55\m$ respectively $150\m$ downstream the IP.
The polarisation measurement is thereby performed at a secondary focus point behind the IP.
This second energy and polarisation measurements enables one to measure and monitor collision effects and gives the possibility to cross-calibrate the individual detectors before and after the IP in the case of ILC runs without colliding beams.

\section{Polarisation Measurement Concept}
\begin{figure}
 \centering
 \includegraphics[width=0.97\textwidth]{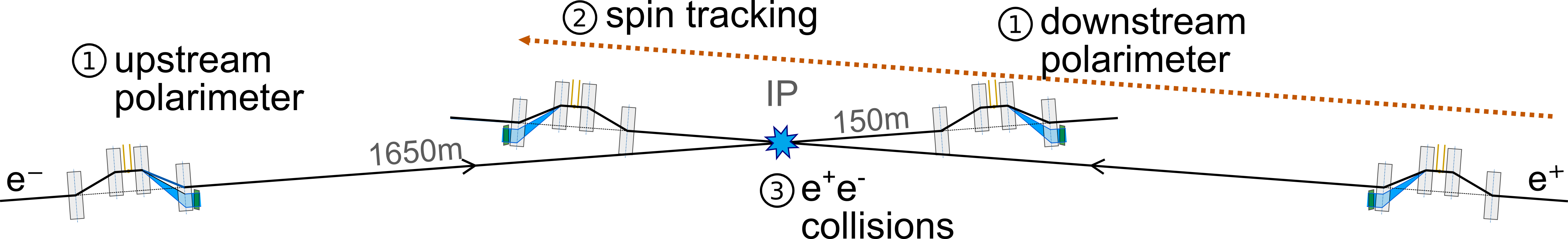}
 \caption{Polarimetry scheme at the ILC: (1) direct up- and downstream polarisation measurement, (2) detailed spin tracking from the polarimeters to the interaction point, and (3) absolute polarisation measurement in $e^+e^-$ annihilation data.}
 \label{fig:polarimetry_scheme}
\end{figure}

The polarisation measurement concepts at the ILC rests on three pillars (c.f. Figure~\ref{fig:polarimetry_scheme}).
Firstly, there is the direct beam polarisation measurement up- and downstream the IP making use of spin dependent scattering processes.
At ILC energies, the process of choice is Compton scattering of laser light off the high energetic leptons.
This way of measuring the polarisation is advantages since it is non-invasive and can be performed in parallel to regular data taking runs.
Secondly, a detailed spin tracking is essential in order to relate the polarisation measurements in the polarimeters to the polarisation at the interaction point.
According to the T-BMT equation~\cite{Thomas:1926dy, Bargmann:1959gz} the effective spin vector of a particle bunch precesses in the presence of magnetic (and electric) fields.
Detailed spin tracking studies can be found in References~\cite{thesisMoritz} and \cite{Beckmann:2014mka} considering also depolarising effects due to beam-beam-interactions.
Thirdly, the absolute value of the beam polarisation can also be determined directly from $e^+e^-$ annihilation data measuring the total and/or differential cross sections of spin dependent processes for different beam polarisation configurations of the electron and positron beam~\cite{thesisIvan, Moenig}.
However, this kind of polarisation measurement is very slow and only after years of data taking a precision comparable to the direct polarimeter measurement can be reached.
Nevertheless, the $e^+e^-$ annihilation data gives indispensably the absolute calibration of the polarimeter measurements.
In the following, we will summarise the challenges and recent achievements in polarimetry at the ILC.

\begin{figure}
 \centering
 \includegraphics[width=0.97\textwidth]{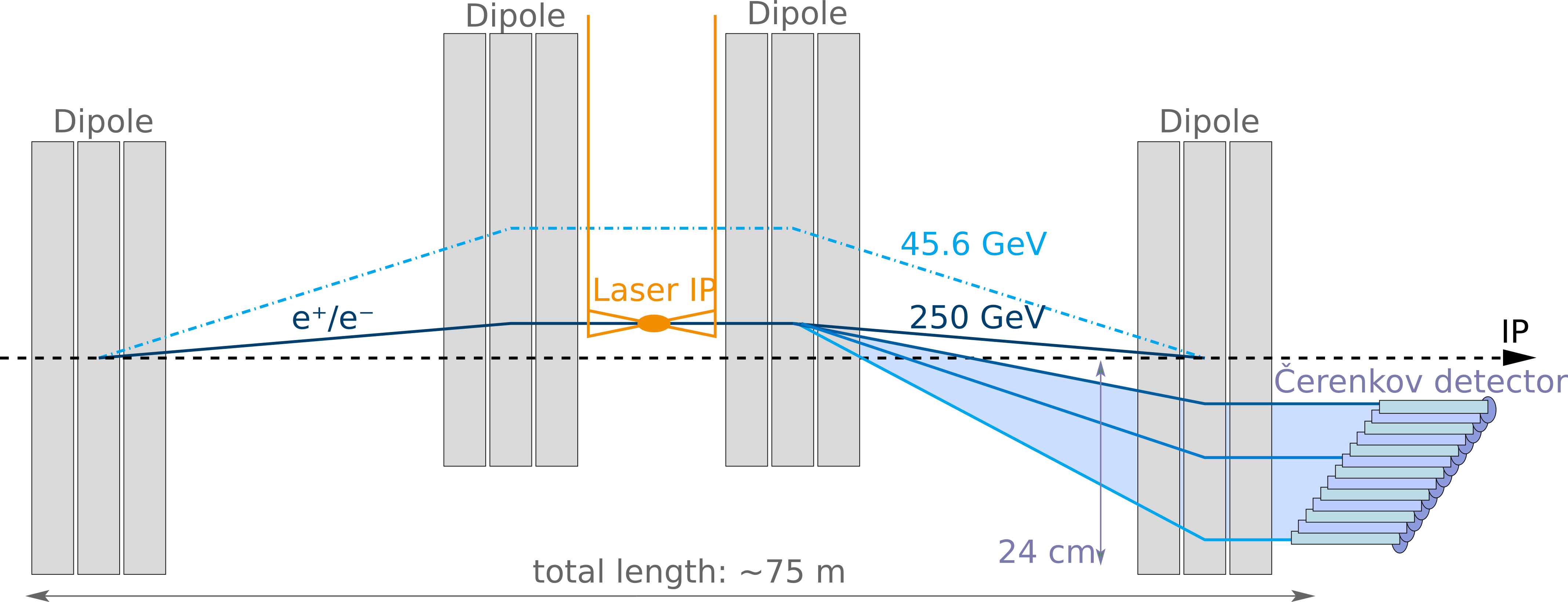}
 \caption{Schematic picture of the upstream polarimeter chicane. Figure taken from Reference~\cite{Boogert:2009ir}.}
 \label{fig:polarimeter_chicane}
\end{figure}

In a magnetic chicane consisting of four dipole magnets (c.f. Figure~\ref{fig:polarimeter_chicane}) the beam is offset by a few centimeters~\cite{Boogert:2009ir}.
A circularly polarised and pulsed laser is shot into the beam under a very small angle.
Out of the whole bunch, in the order of $\mathcal{O}(10^3)$ electrons/positrons are Compton scattered in a very narrow cone.
The energy spectrum of the scattered leptons depends on the product of the laser and the beam polarisation.
This energy distribution is translated into a spatial distribution in the second part of the magnetic chicane.
The Compton scattered leptons are detected by an array of counting Cherenkov detectors next to the beam pipe.
Since the laser can be polarised to almost $100\%$ and the sign of the laser polarisation can be flipped rapidly, the polarisation can be determined from the asymmetry between the cross sections corresponding to the left and right laser helicity
\begin{equation}
 \mathcal{P}=\frac{1}{\mathcal{A}}\frac{\sigma_R-\sigma_L}{\sigma_R+\sigma_L}=\frac{1}{\mathcal{A}}\frac{N^\mathrm{C.e}_R-N^\mathrm{C.e}_L}{N^\mathrm{C.e}_R+N^\mathrm{C.e}_L},\label{eq:polarisation}
\end{equation}
where $\mathcal{A}$ is the so-called analysing power, which is the theoretically calculated asymmetry per detector channel assuming maximal beam and laser polarisation.

At the ILC, a polarisation measurement precision of $\Delta \mathcal{P}/\mathcal{P}=0.25\%$ is envisaged.
The main contributions to the systematic uncertainties are expected to be the laser polarisation, the detector alignment, and the detector nonlinearity.
Thereby, the uncertainty originating from the laser polarisation must not exceed $0.1\%$.
This has already been shown to be feasible at the SLC polarimeter~\cite{Abe:2000dq}.
The detector--beam alignment is essential in order to predict the analysing power correctly and the overall contribution to the error budget must stay below $0.2\%$ translating into an necessary detector alignment at the level of $\mathcal{O}(100\mum)$ and $\mathcal{O}(1\mrad)$.
Finally, the contribution of the detector nonlinearity to the overall error budget must stay below $0.1\%$ in order to meet the precision goal, which means that the detector nonlinearity must be smaller than $0.5\%$.
Additionally, the polarimeter detectors need to be very radiation hard (ionizing dose: $1\MGy$/year for silicon) and fast (readout rate: $1.3\MHz$) since at the upstream polarimeter every bunch of an ILC bunch train is foreseen to be probed.

A natural choice for the detector concept is a gas Cherenkov counter.
A two channel prototype has been developed at DESY based on the gas Cherenkov detector operated at SLC~\cite{Woods:1996nz}.
It consists of a U-shaped, gas-filled aluminum tube with a cross section of $1\times1\cm^2$.
At the end of the horizontal leg, mirrors are installed such that the produced Cherenkov light is reflected upwards to a photomultiplier.
At the end of the other vertical leg, there is an LED light calibration source foreseen.
The prototype has successfully been operated in test beam and it has been shown that asymmetries in the intra-channel light distribution can be utilised in order to align the detector~\cite{Bartels:2010eb}.
Furthermore, it has been demonstrated in a first study that a segmented photomultiplier readout allows to almost reach the mentioned alignment requirements.

\begin{figure}
 \centering
 \includegraphics[width=0.4\textwidth]{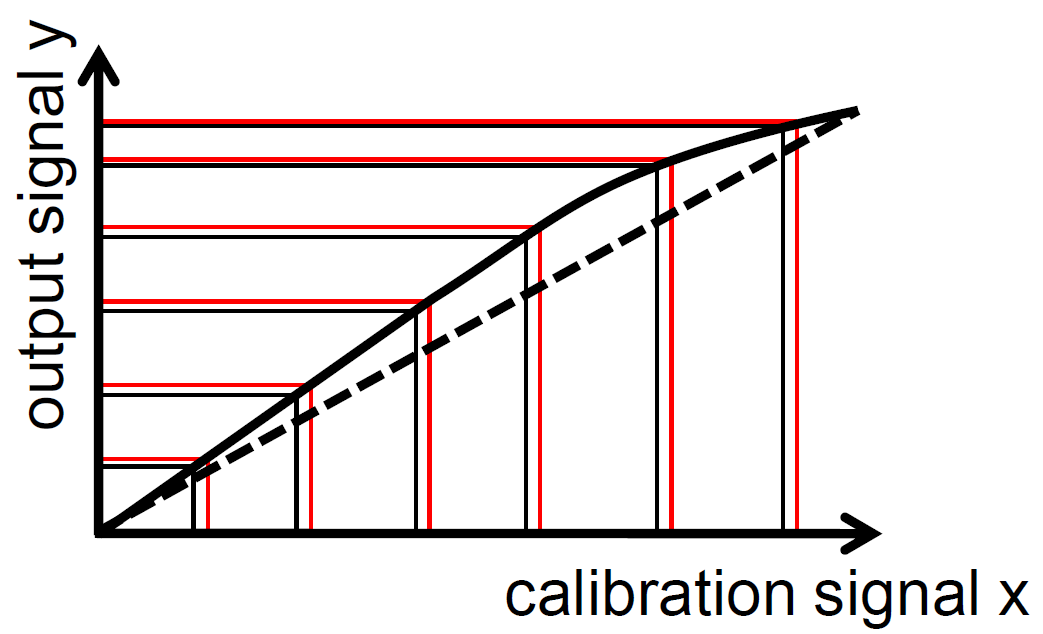}
 \includegraphics[width=0.4\textwidth]{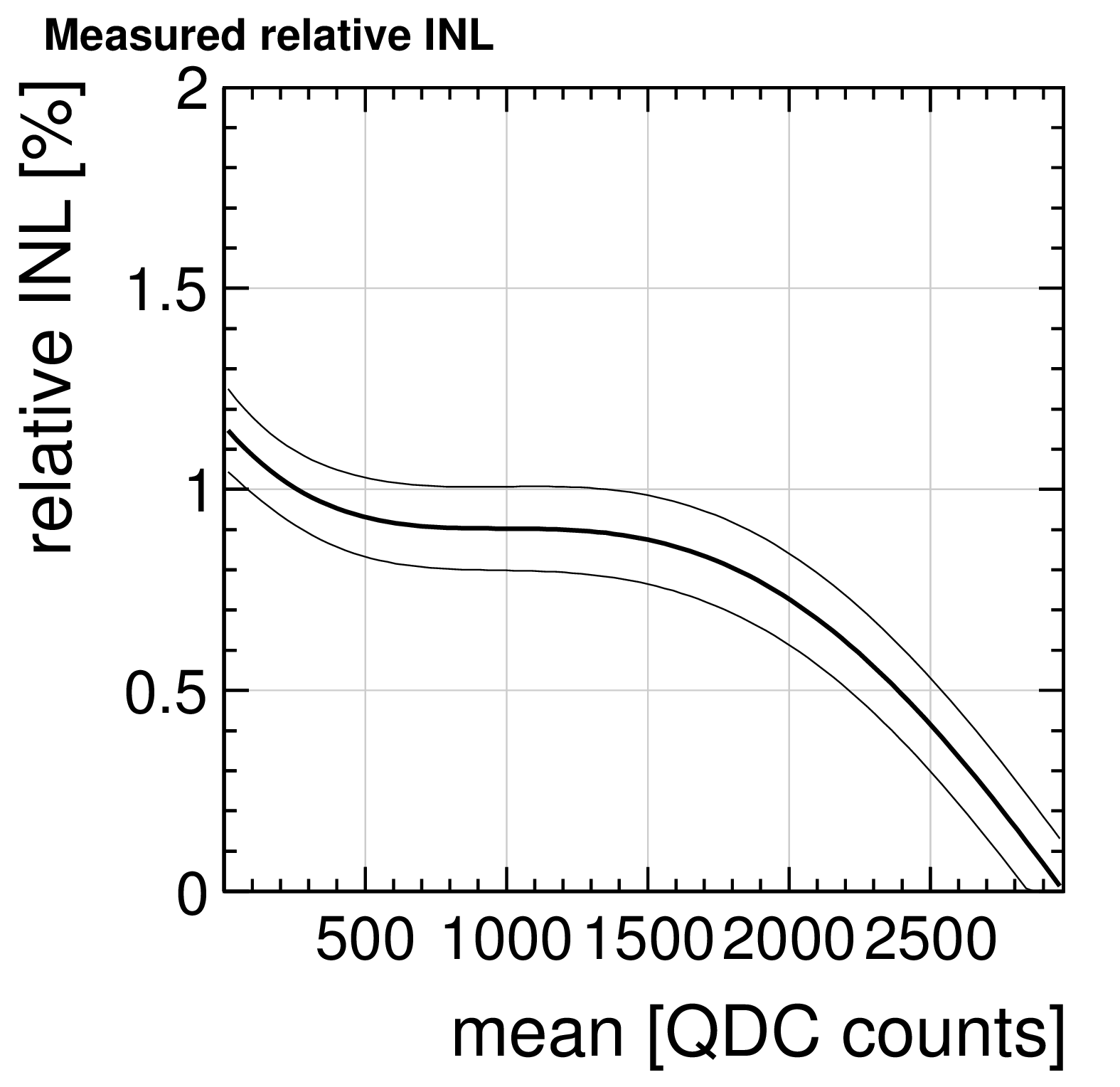}
 \caption{(Left) Illustration of the concept of the proposed differential nonlinearity measurement: two light pulses are used to measure the derivative of the detector transfer function. (Right) Example for the measured relative nonlinearity of a photomultiplier in a dedicated test setup. Figure taken from Reference~\cite{Vormwald:thesis}.}
 \label{fig:dnl_scheme}
\end{figure}

In order to reach the detector linearity design goal, a novel differential calibration method has been developed~\cite{Vormwald:thesis}.
An LED system consisting of two independent LEDs can be used in order to measure the differential nonlinearity of a light detector.
Therefore, one LED light pulse (base pulse $B$) can be scanned over the whole dynamic range of the photodetector.
The second light pulse (differential pulse $D$) is small compared to the base pulse and constant.
The measurement of the difference of the detector response $T(B+D)-T(B)$ returns a measure of the derivative of the detector transfer function $T(x)$ (c.f. Figure \ref{fig:dnl_scheme} (left)).
This measurement can be utilised in order to reconstruct the relative nonlinearity of the detector transfer function to the anticipated precision (c.f. Figure \ref{fig:dnl_scheme} (right)).
Note that according to Equation \eqref{eq:polarisation} no absolute photodetector calibration is necessary. Only a linear correlation between the detector response and the number of primary Compton electrons/ Cherenkov photons needs to be established.
The differential calibration scheme could be operated even during an ILC physics run making use of the gap of about $199\ms$ between two bunch trains.
By this, the whole dynamic range could be calibrated in a floating calibration scheme within about $5\h$.

\begin{figure}
 \centering
 \includegraphics[width=0.5\textwidth]{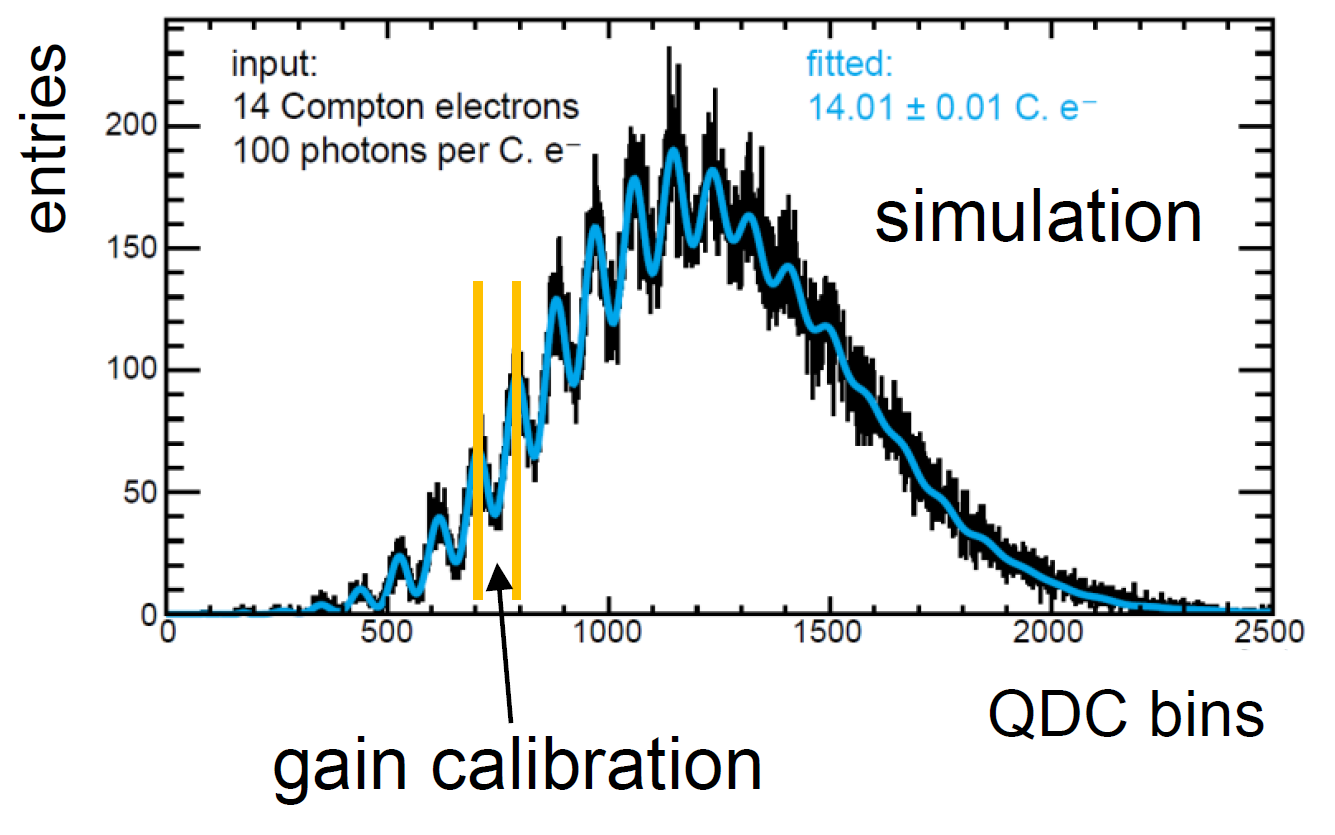}
  \includegraphics[width=0.48\textwidth]{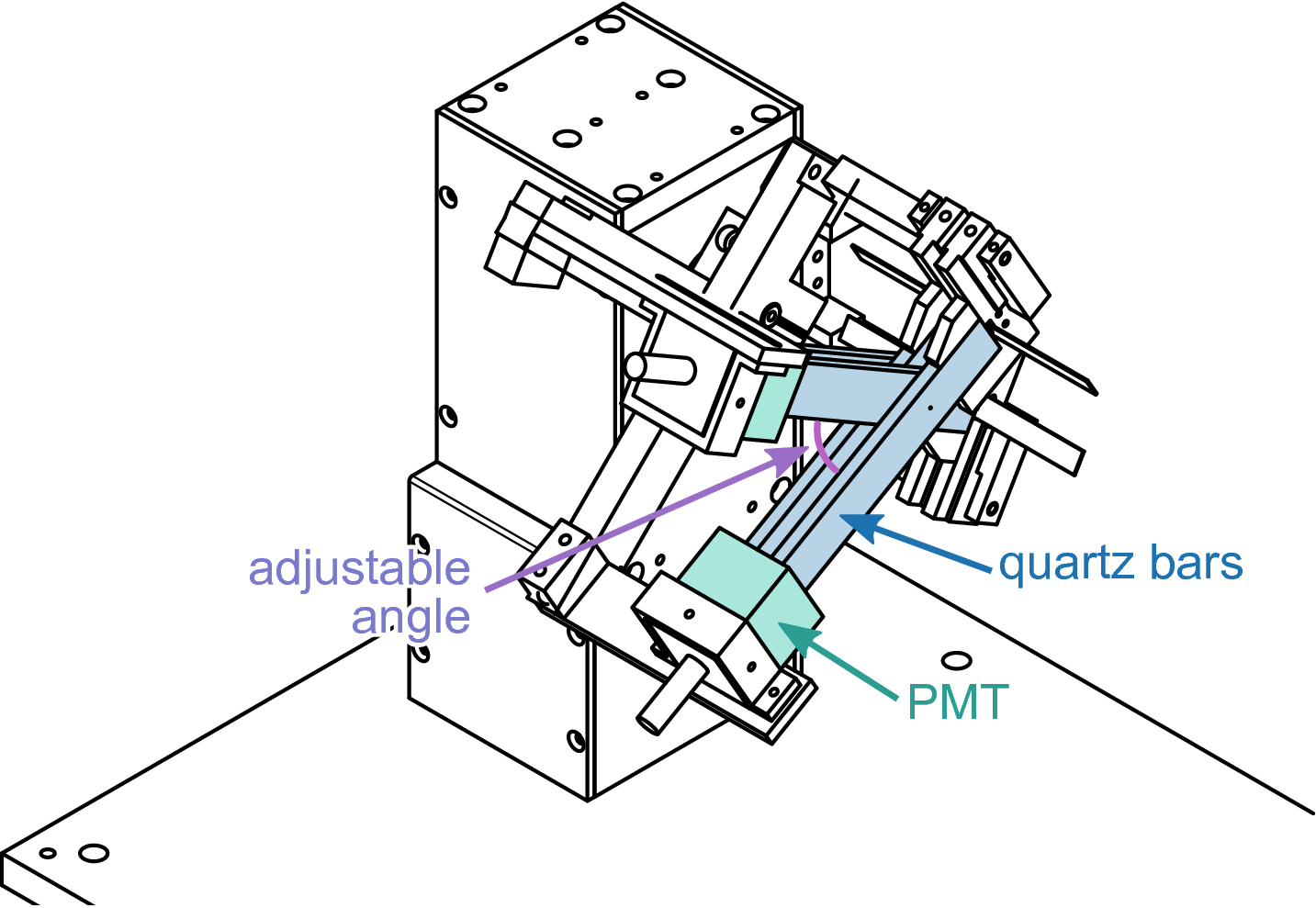}
 \caption{(Left) Simulation of the detector response of a quartz Cherenkov detector. From a fit to the multi-peak structure the average number of initial Compton electrons can be determined very accurately. (Right) Technical drawing of quartz Cherenkov detector prototype. The angle of the quartz bars with respect to the beam axis can be adjusted. Figures taken from Reference~\cite{annika_private}}
 \label{fig:quartz}
\end{figure}

An alternative concept to a gas Cherenkov detector (which needs to be calibrated) is a self-calibrating Cherenkov detector based on quartz~\cite{talkannika}.
Quartz has a much larger refractive index ($n\approx1.45$) compared to the gas foreseen in the gaseous detector ($C_4F_{10}$, $n\approx1.0014$) which results in a much larger light yield per Compton electron.
If the light yield is sufficiently large compared to the number of Compton electrons in a detector channel, it is possible to directly reconstruct the average number of Compton electrons from the detector response (c.f. Figure~\ref{fig:quartz} (right)).
Thus, this detector can be operated without a calibration system.
Furthermore, a combination of a gas and a quartz based Cherenkov detector could be operated behind each other in order to provide additional cross-calibration for the gaseous detector.
It has to be pointed out that the additional light yield of quartz comes at the cost of a much lower energy threshold for Cherenkov light production, which allows to operate this kind of detector only in the clean environment of the upstream polarimeter.

The dimensions of the quartz bars for this concept have been optimised in a Geant4 simulation~\cite{geant4}.
In order to test the whole concept, a four-channel prototype has been built with $5\mm\times18\mm\times100\mm$ quartz bars read out by multianode photomultipliers. 
The incident angle of the electrons in the quartz bar can be adjusted mechanically (c.f. Figure~\ref{fig:quartz} (right)).
The prototype has been successfully operated in the DESY test beam.
Data analysis is still ongoing, but first results look very promising.

\section{Energy Measurement Concept}
\begin{figure}
 \centering
 \includegraphics[width=0.7\textwidth]{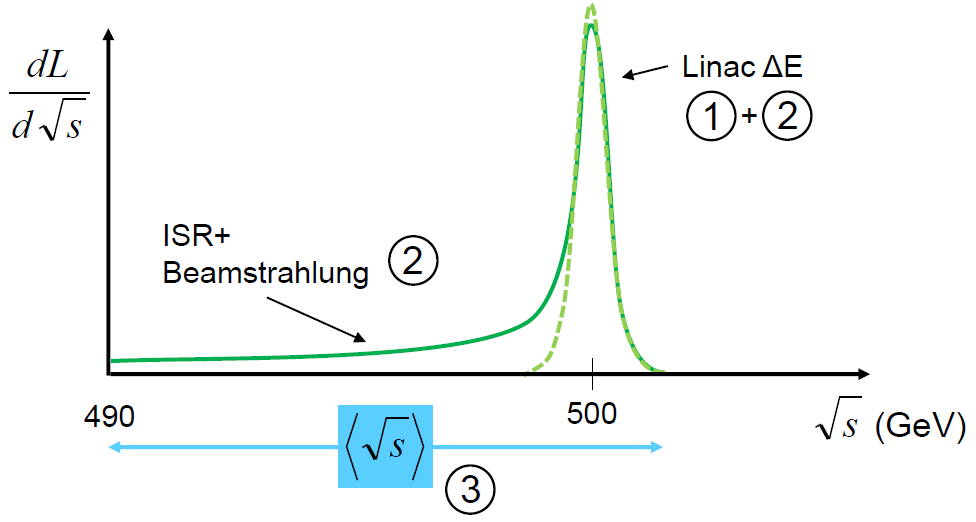}
 \caption{Schematic picture of the beam-energy luminosity spectrum at the ILC before (dashed, light green line) and after (solid, dark green line) collisions. The different foreseen methods for determining the beam energy are sensitive to different parts of the spectrum.}
 \label{fig:energy_scheme}
\end{figure}

Figure \ref{fig:energy_scheme} shows a schematic picture of the luminosity spectrum at the ILC.
The dashed, light green line shows the LINAC beam energy spread before the collision, which is in the order of $\mathcal{O}(0.1\%)$.
After the collision, the luminosity spectrum is broadened towards lower energies due to initial state radiation and beamstrahlung.

At the ILC, three ways of beam energy measurement are foreseen:
An upstream energy spectrometer measures in a non-invasive way the beam energy before the collision (1).
A second energy spectrometer downstream the IP is also sensitive to the energy losses due to collision effects (2).
Finally, the average beam energy can also be determined directly from collision data based on well known processes like radiative $Z$ return events ($\electron^+\electron^-\rightarrow \muon\muon\gamma$) (3)~\cite{Hinze:2005xt}.
In the following, we will focus on the description of the status of the two direct energy measurement concepts (1) and (2).

\begin{figure}
 \centering
 \includegraphics[scale=0.3]{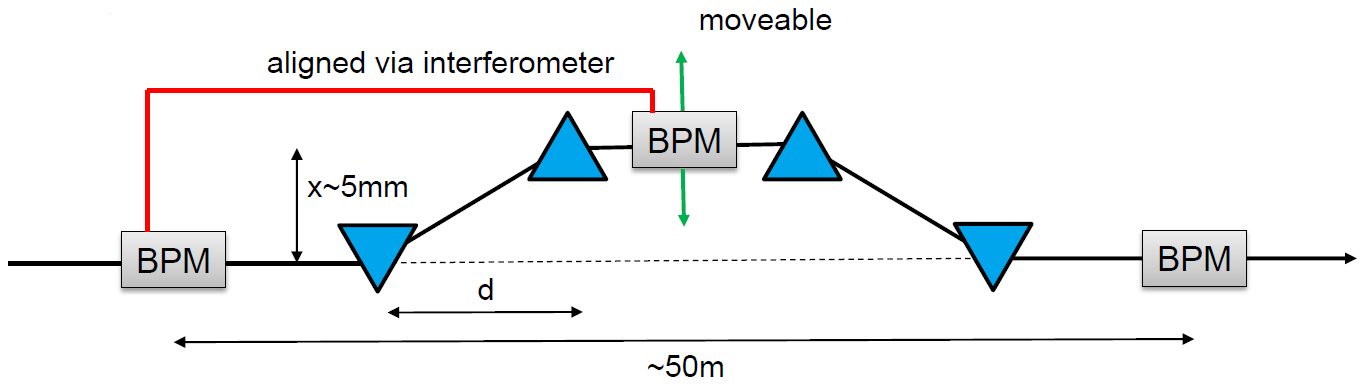}
 \caption{Schematic drawing of the upstream energy spectrometer at the ILC. The beam offset which is induced by a system of dipoles is proportional to the beam energy. A system of interferometrically aligned beam positioning monitors measures the offset.}
 \label{fig:energy_spectrometer_up}
\end{figure}

In the ILC baseline design, the upstream energy spectrometer consists of a magnetic chicane like schematically depicted in  Figure~\ref{fig:energy_spectrometer_up}.
The set of dipole magnets offsets the beam by not more than $5\mm$ in order to keep the induced emittance growth small.
A system of beam positioning monitors (BPMs) which are aligned via an interferometer measures precisely the beam offset $x$.
Assuming that the magnetic field $\vec{B}$ can be mapped sufficiently well along the beam trajectory $\vec{l}$, the beam energy is given by
\begin{equation}
 E_b=q c \frac{d}{x} \int \vec{B}d\vec{l},
\end{equation}
where $d$ is the distance between the first and second dipole magnet of the chicane.
This way of measuring the beam energy is inspired by the energy spectrometer operated at LEP which demonstrated to be capable to measure the beam energy to a precision of $\Delta E/E=0.019\%$~\cite{Assmann:2004gc}.
In the LEP energy spectrometer, only one dipole magnet was used and the beam bending angle was measured.
In order to minimise the beam inclination in the BPMs and, thus, maximise the precision of the measurement, the specific chicane setup has however been chosen for the ILC baseline design.
The envisaged beam energy measurement precision amounts to $\mathcal{O}(0.1\%)$ over the whole ILC energy range ($45.6\GeV-500\GeV$), which requires a resolution in $x$ of better than $0.5\mum$~\cite{Boogert:2009ir}.

It is clear that in order to meet this precision goal, ultra-precise BPMs are necessary. 
Recent developments of cavity BPMs can meet these requirements.
In test beam studies in the ATF extraction line at KEK, it has been demonstrated that a position resolution of $15.6\nm$ and a tilt resolution of $2.1\murad$ of a BPM is achievable~\cite{Walston:2007rj}.
  
Furthermore, a complete ILC-like energy chicane with a set of BPMs and a dispersion of $x=5\mm$ has been set up in the End Station A (ESA) at SLAC~\cite{Slater:2008zz}.
Thereby, the available beam parameters were very comparable to the ILC design beam parameters in terms of repetition rate, bunch charge, bunch length, and relative energy spread.
The setup allowed for stable measurements on $\mathcal{O}(1\mum)$ over one hour and a resolution of the beam orbit reconstruction of $0.8\mum/1.2\mum$ in $x/y$ direction has been reported~\cite{Slater:2008zz}. In a further study, an energy resolution for a single bunch measurement of $\Delta E/E=0.05\%$ has been determined and several issues for further improvement of the setup were identified~\cite{Lyapin:2010sb}. Thus, the targeted precision seems to be in reach.

For the downstream energy spectrometer, a different concept is foreseen~\cite{eric}.
In a magnetic chicane consisting of three horizontal bending magnets the beam is deflected.
In between the dipoles, vertical wigglers are placed.
Therefore, in addition to a horizontal synchrotron radiation fan caused by the  horizontal deflection in the chicane, two perpendicular synchrotron strips are produced by the wigglers.
The opening angle of those two strips, which is in the order of $4\mrad$, is a measure for the beam energy.
In a distance of $100\m$ from the analysing (middle) dipole, the distance of the synchrotron strips is determined in a detector plane.
Vertically well-aligned quartz fibers with a diameter of $100\mum$ act as scintillator and Cherenkov medium for converted electrons/positrons from synchrotron radiation photons.
Those fibers are read out by photomultipliers.
The measurement of the distance between the strips below and above the horizontal synchrotron radiation fan allows in addition to control systematic effects in the alignment.
Due to the rather large beam energy spread downstream of the IP, the synchrotron radiation strip is distributed over more than one fibre.
Thus, in this method not only the peak energy is accessible, but also energy losses in the collisions can be monitored.
A similar detector was already operated at SLC using wires instead of quartz fibres~\cite{Kent:1990zy}.
An energy resolution of $\Delta E/E=0.02\%$ has been achieved there.
The precision goal at the ILC amounts to $\mathcal{O}(0.01\%)$, which requires a fibre alignment to $20\mum$.
A prototype consisting of 60 active fibres spaced on a pitch of $200\mum$ has been build and successfully operated in ESA at SLAC~\cite{eric}.

\begin{figure}
 \centering
 \includegraphics[scale=0.3]{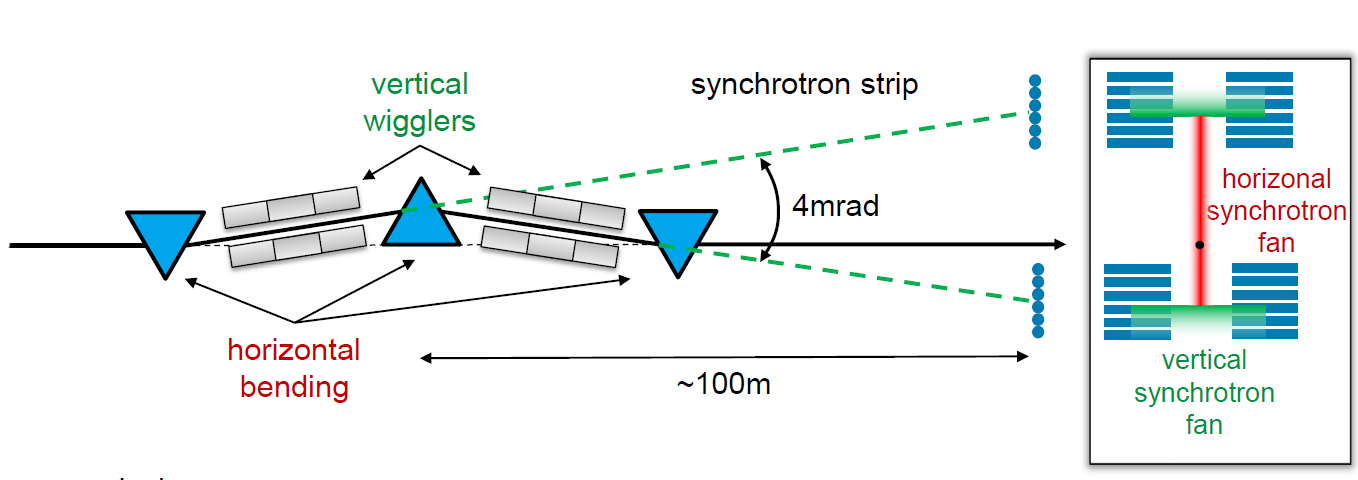}
 \caption{Schematic drawing of the downstream energy spectrometer at the ILC. A set of wigglers produce two vertical synchrotron radiation strips whose opening angle is a measure for the beam energy. A quartz fiber detector situated in a distance of $100\m$ measures the distance between the two strips. This detection method is also sensitive to the beam energy losses in the collision.}
 \label{fig:energy_spectrometer_down}
\end{figure}

\section{Conclusions}
Excellent beam instrumentation plays a crucial role in the precision physics programme of the ILC.
The beam polarisation as well as the beam energy are important quantities in many analyses.
For the polarisation measurement, two different polarimeter detector concepts based on quartz and gas have been developed, which are applicable in different beam line positions and complement each other in terms of cross-calibration.
The energy measurement is performed in two independent ways before and after the interaction point. Thereby, both, the peak beam energy and energy losses in the collisions are accessible.

Concluding, the polarisation as well as the energy measurement concepts at the ILC are in a very good shape.
For all the concepts it could be shown on the prototype level that the challenging precision goals are in reach and no major obstacles are in sight.

\acknowledgments

The author would like to thank Jenny List, Annika Vauth, Moritz Beckmann, and Eric Torrence for helpful discussions and valuable input.


\begin{thebibliography}{9}

\bibitem{tdr}
   ILC Technical Design Report, 2013,
   Volume 1: \reftitle{Executive Summary} \href{http://arxiv.org/abs/arXiv:1306.6327}{arXiv:1306.6327v1 [physics.ins-det]},
   Volume 2: \reftitle{Physics} \href{http://arxiv.org/abs/arXiv:1306.6352}{arXiv:1306.6352v1 [physics.ins-det]},
   Volume 3.I:  \reftitle{Accelerator R\&D} \href{http://arxiv.org/abs/arXiv:1306.6353}{arXiv:1306.6353v1 [physics.acc-ph]},
   Volume 3.II: \reftitle{Accelerator Baseline Design} \href{http://arxiv.org/abs/arXiv:1306.6328}{arXiv:1306.6328v1 [physics.acc-ph]},
   Volume 4: \reftitle{Detectors} \href{http://arxiv.org/abs/arXiv:1306.6329}{arXiv:1306.6329v1 [physics.ins-det]}.

\bibitem{Thomas:1926dy}
  L.~H.~Thomas,
  \reftitle{The motion of a spinning electron}
  Nature {\bf 117} (1926) 514.

\bibitem{Bargmann:1959gz}
  V.~Bargmann, L.~Michel and V.~L.~Telegdi,
  \reftitle{Precession of the polarization of particles moving in a homogeneous electromagnetic field}
  Phys.\ Rev.\ Lett.\  {\bf 2} (1959) 435.

\bibitem{thesisMoritz}
   M.~Beckmann,
   \reftitle{Spin Transport at the International Linear Collider and its Impact on the Measurement of Polarization}
   Ph.~D. Thesis, Hamburg University,
   2013,
   \href{http://www-library.desy.de/cgi-bin/showprep.pl?desy-thesis-13-053}{DESY-THESIS-13-053}.

\bibitem{Beckmann:2014mka}
  M.~Beckmann, J.~List, A.~Vauth and B.~Vormwald,
  \reftitle{Spin Transport and Polarimetry in the Beam Delivery System of the International Linear Collider}
  arXiv:1405.2156 [physics.acc-ph].

\bibitem{thesisIvan}
   I.~Marchesini,
   \reftitle{Triple Gauge Couplings and Polarization at the ILC and Leakage in a Highly Granular Calorimeter}
   Ph.~D. Thesis, Hamburg University,
   2011,
   \href{http://www-library.desy.de/cgi-bin/showprep.pl?desy-thesis-11-044}{DESY-THESIS-11-044}.

\bibitem{Moenig}
  K.~M\"onig,
  ``Polarisation Measurements With Annihilation Data,''
  LC-PHSM-2004-012.

\bibitem{Boogert:2009ir}
  S.~Boogert, M.~Hildreth, D.~Kafer, J.~List, K.~Monig, K.~C.~Moffeit, G.~Moortgat-Pick and S.~Riemann {\it et al.},
  \reftitle{Polarimeters and Energy Spectrometers for the ILC Beam Delivery System}
  JINST {\bf 4} (2009) P10015,
  arXiv:0904.0122 [physics.ins-det].

\bibitem{Abe:2000dq}
  K.~Abe {\it et al.}  [SLD Collaboration],
  \reftitle{A High precision measurement of the left-right Z boson cross-section asymmetry}
  Phys.\ Rev.\ Lett.\  {\bf 84} (2000) 5945,
  arXiv:hep-ex/0004026.

\bibitem{Woods:1996nz}
  M.~Woods [SLD Collaboration],
  \reftitle{The Scanning Compton polarimeter for the SLD experiment}
  arXiv:hep-ex/9611005.

\bibitem{Bartels:2010eb}
  C.~Bartels, J.~Ebert, A.~Hartin, C.~Helebrant, D.~Kafer and J.~List,
  \reftitle{Design and Construction of a Cherenkov Detector for Compton Polarimetry at the ILC}
  JINST {\bf 7} (2012) P01019,
  arXiv:1011.6314 [physics.ins-det].

\bibitem{Vormwald:thesis}
   B.~Vormwald, 
   \reftitle{From Neutrino Physics to Beam Polarisation -- a High Precision Story at the ILC} 
   Ph.~D. Thesis, Hamburg University,
   2014,
   \href{http://www-library.desy.de/cgi-bin/showprep.pl?desy-thesis-14-006}{DESY-THESIS-14-006}.

\bibitem{annika_private}
   A.~Vauth, private communication, 2014.

\bibitem{talkannika}
  A.~Vauth,
  \reftitle{A Quartz Cherenkov Detector for Polarimetry at the ILC} 
  EUCARD workshop "Spin optimization at Lepton accelerators", Mainz, February 12 - 13, 2014,
  \href{https://indico.mitp.uni-mainz.de/contributionDisplay.py?contribId=6&sessionId=11&confId=18}{https://indico.mitp.uni-mainz.de/contributionDisplay.py?contribId=6\&sessionId=11\&confId=18}.

\bibitem{geant4}
  S.~Agostinelli {\it et al.}  [GEANT4 Collaboration],
  \reftitle{GEANT4: A Simulation toolkit}
  Nucl.\ Instrum.\ Meth.\ A {\bf 506} (2003) 250.
  
  J.~Allison, K.~Amako, J.~Apostolakis, H.~Araujo, P.~A.~Dubois, M.~Asai, G.~Barrand and R.~Capra {\it et al.},
  \reftitle{Geant4 developments and applications}
  IEEE Trans.\ Nucl.\ Sci.\  {\bf 53} (2006) 270.

\bibitem{Hinze:2005xt}
  A.~Hinze,
  \reftitle{Determination of beam energy at TESLA using radiative return events}
  LC-PHSM-2005-001.

\bibitem{Assmann:2004gc}
  R.~Assmann {\it et al.}  [LEP Energy Working Group Collaboration],
  \reftitle{Calibration of centre-of-mass energies at LEP 2 for a precise measurement of the W boson mass}
  Eur.\ Phys.\ J.\ C {\bf 39} (2005) 253,
  arXiv:hep-ex/0410026.

\bibitem{Walston:2007rj}
  S.~Walston, S.~Boogert, C.~Chung, P.~Fitsos, J.~Frisch, J.~Gronberg, H.~Hayano and Y.~Honda {\it et al.},
  \reftitle{Performance of a High Resolution Cavity Beam Position Monitor System}
  Nucl.\ Instrum.\ Meth.\ A {\bf 578} (2007) 1.

\bibitem{Slater:2008zz}
  M.~Slater, C.~Adolphsen, R.~Arnold, S.~Boogert, G.~Boorman, F.~Gournaris, M.~Hildreth and C.~Hlaing {\it et al.},
  \reftitle{Cavity BPM system tests for the ILC energy spectrometer}
  Nucl.\ Instrum.\ Meth.\ A {\bf 592} (2008) 201.

\bibitem{Lyapin:2010sb}
  A.~Lyapin, H.~J.~Schreiber, M.~Viti, C.~Adolphsen, R.~Arnold, S.~Boogert, G.~Boorman and M.~V.~Chistiakova {\it et al.},
  \reftitle{Results from a prototype chicane-based energy spectrometer for a linear collider}
  JINST {\bf 6} (2011) P02002,
  arXiv:1011.0337 [physics.acc-ph].

\bibitem{eric}
  E.~Torrence,
  \reftitle{Extraction Line Energy Spectrometer}
  Linear Collider Workshop of the Americas (ALCPG11), Eugene, March 19 -23, 2011,
  \href{http://agenda.linearcollider.org/contributionDisplay.py?contribId=186&sessionId=17&confId=4572}{http://agenda.linearcollider.org/contributionDisplay.py?contribId=186\&sessionId=17\&confId=4572}.

\bibitem{Kent:1990zy}
  J.~Kent, J.~J.~Gomez Cadenas, A.~Hogan, M.~King, W.~Rowe, S.~Watson, C.~Von Zanthier and D.~Briggs {\it et al.},
  \reftitle{Design Of A Wire Imaging Synchrotron Radiation Detector}
  SLAC-PUB-5110.

\end{thebibliography}
\end{document}